\documentstyle[epsfig]{aipproc}

\def\gsim{\mathrel{\rlap{\raise 1.5pt \hbox{$>$}}\lower 3.5pt
\hbox{$\sim$}}}
\def\lsim{\mathrel{\rlap{\raise 2.5pt \hbox{$<$}}\lower 2.5pt
\hbox{$\sim$}}}

\begin{document}

\title{Standard-Model-like scenarios in the 2HDM
and Photon Collider potential}

\author{Ilya F. Ginzburg$^{a}$,
Maria Krawczyk$^{b}$ {\rm and}
Per Osland$^{c}$}

\address{
$^a$ Sobolev Institute of Mathematics, SB RAS,
     630090 Novosibirsk, Russia\\
$^b$ Institute of Theoretical Physics, Warsaw University, Poland\\
$^c$ Department of Physics, University of Bergen,
     Allegt.\ 55, N-5007 Bergen, Norway }

\maketitle

\begin{abstract}
After operations at the LHC and $e^+e^-$ Linear Colliders it may be found 
that a Standard-Model-like scenario is realized. 
In this scenario no new particle will be
discovered, except a single Higgs boson having partial widths or
coupling constants with fundamental particles, whose squares are close,
within anticipated experimental uncertainty, to those of the SM. 
Experiments at a Photon Collider can resolve whether the SM
or e.g.\ the Two-Higgs-Doublet Model is realized in Nature.

For the SM-like realizations  of the 2HDM~(II) we study 
the loop couplings of 
the Higgs boson with $\gamma\gamma$ and $Z \gamma$, and also with gluons. 
The deviation of the two-photon width from its SM value 
is generally higher than the expected inaccuracy 
in the measurement of $\Gamma_{\gamma \gamma}$ at a Photon Collider. 
The deviation is sensitive to the parameters of the Higgs self interaction.
\end{abstract}

\section{Introduction. SM-like scenario}

It could happen that no new particles will be discovered at
the Tevatron, the LHC and $e^+e^-$ Linear Colliders 
\cite{Accomando:1998wt} except the SM-like Higgs boson. 
In this case the main task for new colliders 
will be to search for signals of new physics via deviations of
observed quantities from Standard-Model predictions.  The study
of Higgs boson production at a Photon Collider \cite{GKST}
offers excellent opportunities for this \cite{Ginzburg:1999fb}. 
Indeed, in the SM and in its extensions, all
fundamental charged particles contribute to the $h\gamma \gamma$
and $hZ\gamma$ effective couplings.  Besides, these couplings
are absent in the SM at tree level, appearing only at the loop
level. Therefore, the background for signals of new physics will
be relatively lower here than in processes which are
allowed at tree level of the SM.

We assume \cite{gko} that an SM-like scenario is realized,
i.e., the Higgs particle has been found at the Tevatron or the
LHC and its partial widths or coupling constants squared are
precisely measured (mainly at the $e^+e^-$ Linear Collider),
being in agreement with those of the SM, within the anticipated
experimental accuracies. This can happen not only in the SM, but also
if Nature is described by some other theory, for example,
the Two-Higgs-Doublet Model (2HDM) or the Minimal Supersymmetric
Standard Model (MSSM). 
Here  we compare the SM and the SM-like scenario in the 2HDM~(II).
\smallskip

The {\it SM-like scenario} can be defined by the following criteria:
\begin{itemize}
\item
One Higgs boson will be discovered with mass above today's limit 
for an SM Higgs boson, 115~GeV \cite{LEPC}. 
This can be either the Higgs boson of the SM or 
one neutral Higgs boson from some other model.
\item  No other Higgs boson will be discovered.
Any other Higgs boson is weakly coupled with the $Z$ boson, 
gluons and quarks, or sufficiently heavy to escape observation:
$M_H,\, M_A,\,
M_{H^{\pm}}> {\cal O}(800\mbox{ GeV})$ \cite{lhc}.
\item
Any other new particle that may exist is beyond
the discovery limits of LHC and the $e^+e^-$ Linear Collider.
\item The measured decay widths of this Higgs boson (or
coupling constants squared) to quarks, charged leptons, 
electroweak gauge bosons and gluons, $\Gamma_i^{\rm exp}$ 
($i=q,l,W,Z,g$), will be in agreement with their SM values 
$\Gamma_i^{\rm SM}$ within the experimental precision,
$|\Gamma_i^{\rm exp}/\Gamma_i^{\rm SM}-1|\ll 1$.
\end{itemize}

\section{Precision of measured SM Higgs couplings}

At the TESLA $e^+e^-$ collider the
discussed production cross sections are expected to be measured
with a significantly higher precision than at the LHC \cite{lhc}.
At $M_h=120~\mbox{GeV}$, with integrated luminosity 500~fb$^{-1}$
one can expect for the basic couplings the following relative 
accuracy \cite{Battaglia:2001jb}:
\begin{equation}
\delta_b=0.021, \qquad  \delta_t=0.022, \qquad
\delta_{W/Z}=0.012.
\label{couplacc}
\end{equation}
Experiments at Photon Colliders open new perspectives. 
Even with a modest integrated luminosity of
a $\gamma\gamma$ collider in the high energy peak
of about $40$~fb$^{-1}$, a $\gamma\gamma$ collider
makes it possible to obtain the accuracy in measuring 
$h\gamma\gamma$  \cite{Jikia:2000en}:
\begin{equation}
\delta_\gamma= 0.02\quad \mbox{ for }M_h<140\mbox{ GeV},
\label{jik}
\end{equation}
to be compared to 13\% at a linear collider at 500~fb$^{-1}$ 
\cite{Battaglia:2001jb,boos}.
The accuracy in the measurement of the effective $hZ\gamma$
($HZ\gamma$) coupling in the process $e\gamma\to eh$
($e\gamma\to eH$) is lower.

We will use the above uncertainties to constrain  ratios of
actual (in principle measurable) coupling constants of each neutral
Higgs scalar $\phi$ ($h$ or $H$) of the 2HDM (II) in the SM-like 
realizations\footnote{Discussing both these scalars, 
we use the notation $\phi$ for $h$ and $H$.}
with particle $i$ 
to the corresponding value for the Higgs boson in the SM,
\begin{equation}
\chi_i^\phi= \frac{g_i^\phi}{g_i^{\rm SM}},\quad {\rm where}\quad
\chi_i^{\phi}=\pm (1-\epsilon_i),\quad
\mbox{with }  |\epsilon_i|\ll 1\,.
\label{estacc}
\end{equation}

The allowed ranges for $\epsilon_i$ are constrained by the
experimental accuracies $\delta_i$, $|\epsilon_i|\le \delta_i$. 
Additional constraints follow from the structure of the considered 
model.

\section{Two-Higgs-Doublet Model (II)}

We here consider the CP-conserving Two-Higgs-Doublet Model in its 
Model~II implementation, denoted by
2HDM~(II) \cite{Hunter,Santos:1997vt,Haber:1995be}. Here, one doublet of
fundamental scalar fields couples to $u$-type quarks, the
other to $d$-type quarks and charged leptons. The Higgs sector
contains three neutral Higgs particles, two CP-even scalars $h$ and
$H$, one CP-odd (pseudoscalar) $A$, and charged Higgs bosons
$H^\pm$ --- it coincides in the 2HDM~(II) and in the MSSM.

In the SM-like scenario realized in the 2HDM we need to consider
both possibilities: not only the light scalar Higgs boson, $h$,
but also the heavier one, $H$, could imitate the SM Higgs boson if the
lighter scalar $h$ escapes detection 
\cite{light-Higgs,Grzadkowski:1999ye}.

The ratios of the direct coupling constants of the Higgs bosons
$\phi=h,H$ to the gauge bosons $V=W$ or $Z$ bosons, to up and 
down quarks and to charged leptons, relative to their SM values 
can be expressed via angles $\alpha$ and $\beta$ 
\cite{Hunter,Haber:1995be}:
\begin{equation}
\begin{array}{l}
\chi_V^\phi=\sin(\beta-\alpha+\delta_\phi), \\
\chi_u^\phi=\sin(\beta-\alpha+\delta_\phi)
+\cot\beta\cos(\beta-\alpha+\delta_\phi), \\ 
\chi_d^\phi=\sin(\beta-\alpha+\delta_\phi)
-\tan\beta\cos(\beta-\alpha+\delta_\phi),
\end{array} \qquad
\delta_h=0, \quad  \delta_H=\frac{\pi}{2}\,.
\label{2hdmcoup-h}
\end{equation}
Here, $\beta\in (0,\,\pi/2)$ parameterizes the ratio of the vacuum 
expectation values of the two basic Higgs doublets and 
$\alpha\in(-\pi,\;0)$ parameterizes
mixing among the two neutral CP-even Higgs fields. 

The coupling of the charged Higgs boson to the
neutral scalars $\phi$ depends on the Higgs-boson masses and on
the additional parameter $\lambda_5$ (compare
\cite{Hunter,Djouadi:1998yq}, note that the paper \cite{Haber:1995be} uses
another parameterization of the 2HDM)
\begin{equation}
\chi_{H^\pm}^\phi
=\left(1-\frac{M_\phi^2}{2M_{H^\pm}^2}\right)\chi_V^{\phi}
+\left(\frac{M_\phi^2}{2M_{H^\pm}^2}-\frac{\lambda_5}{2\lambda_4}\right)
(\chi_u^{\phi}+\chi_d^{\phi}), \quad
\lambda_4=\frac{2M_{H^\pm}^2}{v^2}. \label{b2d2}
\end{equation}

\section{ Pattern relation and allowed ranges for couplings }

The quantities $\chi_i^\phi$ for the couplings (\ref{2hdmcoup-h}) of 
each scalar (referred to below as basic couplings)
are closely related to the observables and it is more natural 
to use them, instead of $\alpha$ and $\beta$.
Since for each $\phi$ these three $\chi_i$ can be expressed in
terms of {\it two} angles, they fulfill a simple relation
{\em(pattern relation)}, which plays a basic role in our analysis 
(it is valid in the MSSM as well). 
It has the same form for  both $h$ and $H$, namely $(\chi_u -\chi_V)
(\chi_V -\chi_d) +\chi_V^2=1$, or
\begin{equation}
(\chi_u +\chi_d)\chi_V=1+\chi_u \chi_d.
\label{2hdmrel}
\end{equation}
Furthermore, from Eq.~(\ref{2hdmcoup-h}) follows the condition:
$\tan^2\beta=(\chi_V-\chi_d)/(\chi_u-\chi_V)
=(1-\chi_d^2)/(\chi_u^2-1)$.

In the following discussion we will assume only one value for each
up-type quark, down-type quark, charged lepton and gauge boson coupling
with the Higgs boson, in numerical calculation we will use the best 
estimate for each category, e.g. $\delta_b$ for $\delta_d$.
The SM-like scenario means, in particular, that 
$\chi_i^2\approx 1$ for the basic couplings.
We consider solutions of the equations (\ref{estacc})
constrained by the pattern relation (\ref{2hdmrel}).

These solutions can be further classified as follows. 
For solutions $A_{\phi\pm}$ the relative couplings are
approximately identical, $\chi_V\approx \chi_u\approx
\chi_d\approx \pm 1$.  There are also solutions $B_{\phi \pm q}$,
where some of the $\chi_i\approx 1$ but other $\chi_j\approx -1$.
Here, the first subscript labels the observed Higgs boson
($\phi=h,H$). The second subscript $\pm$ labels the sign of
$\chi_V^\phi$. The third subscript $q=d,u$ (only for solutions
$B$) labels the quark whose coupling constant to the Higgs boson is
opposite to that of the vector boson.
Our analysis \cite{gko} shows that an SM-like scenario can be 
realized for only a limited part of the 2HDM parameter space,
summarized in Table~1.
\begin{table}[htb]
\begin{center}
\begin{minipage}{150mm}
\begin{tabular}{||c|c|c|c|c|c||}
\noalign{\vspace{-8.5pt}} \hline\hline
&observed&&\multicolumn{2}{|c|}{}&\\
& Higgs &$\chi_V$&\multicolumn{2}{|c|}{$\tan\beta$}&constraint\\
&boson&&\multicolumn{2}{|c|}{}&\\ \hline
&&&&$>1$&\\
$A_{\phi\pm}:$
& h&$\approx+1$&&$<1$&\\ 
\cline{2-3}\cline{5-5}
$\chi_v\approx\chi_u\approx\chi_d$& h&$\approx -1$
&$\sqrt{\left|\frac{\epsilon_d}{\epsilon_u}\right|}$ 
& $\gg 1$
&$\epsilon_V=-\frac{\epsilon_u\epsilon_d}{2}$\\
\cline{2-3}\cline{5-5}
&H&$\approx - 1$&&$>1$&\\\hline
& & &\multicolumn{2}{|c|}{} &\\
$B_{\phi\pm d}:$
&h&$\approx +1$
&\multicolumn{2}{|c|}{}&\\\cline{2-3}
$\chi_V\approx \chi_u\approx -\chi_d$  & h&$\approx -1$&
\multicolumn{2}{|c|}{
$\sqrt{\frac{2}{\epsilon_V}}>1$}
& $\epsilon_u=-\frac{\epsilon_V\epsilon_d}{2}$\\
\cline{2-3}
& H&$\approx -1$&\multicolumn{2}{|c|}{}&
\\ \hline
& & &\multicolumn{2}{|c|}{} &\\
$B_{h+u}:$
&h&$\approx+1$&\multicolumn{2}{|c|}{
$\sqrt{\frac{\epsilon_V}{2}}<1$}&
$\epsilon_d=-\frac{\epsilon_V\epsilon_u}{2}$\\
$\chi_V\approx
\chi_d\approx-\chi_u$&&&\multicolumn{2}{|c|}{}&\\ \hline
\multicolumn{6}{||c||}{}\\
\multicolumn{6}{||c||}{
$\chi_i=\frac{g_i}{g_i^{SM}}=\pm(1-\epsilon_i)$ with}\\
\multicolumn{6}{||c||}{
$i=V(\equiv
Z,\,W)$ or $i=u(\equiv t,\, c)$ or $i=d,\ell(\equiv b,\, \tau)$.}\\
\multicolumn{6}{||c||}{
$\epsilon_V>0$, $\epsilon_u\epsilon_d<0$}\\\hline\hline
\end{tabular}
\end{minipage}
\end{center}
\caption{Realizations of SM-like scenario in the 2HDM~(II).}
\end{table}

\section{Distinguishing models via loop couplings}

In order to distinguish models in the considered SM-like scenario,
we compute loop-induced couplings of the Higgs boson with photons
or gluons \cite{Hunter,Djouadi:1998yq} for these solutions $A$ and $B$  
within the ranges of the coupling constants allowed by the anticipated
experimental inaccuracies from Eq.~(\ref{couplacc}). To estimate
the deviation from the SM, we consider the ratios of widths
$|\chi_{\gamma\gamma}|^2$ and $|\chi_{Z \gamma}|^2$ obtained in the
2HDM~(II) and in the SM. In the 2HDM the couplings with photons,
$\phi\gamma\gamma$ and $\phi Z\gamma$, contain contributions from
fermions, and from the charged gauge boson $W^{\pm}$, like in the
SM. In addition, there are contributions from the charged Higgs
boson, $H^{\pm}$. For definiteness, we perform all calculations
for $M_{H^\pm}=800$~GeV. At $M_\phi<250$~GeV the contribution
of the charged Higgs boson loop varies by less than 5\% when
$M_{H^{\pm}}$ varies from 800~GeV to infinity. 
We have kept track of all relevant couplings to make sure they
are in the perturbative regime for observed Higgs boson 
masses up to 3~TeV. 
If the SM-like Higgs is $h$ then any non-perturbative
coupling would correspond to the unobserved Higgs boson $H$.

The $\gamma\gamma$ width looks the most promising one for
distinguishing models. 

{\it Solutions A}. A new feature of the two-photon width, as compared
to the SM case, is the contribution due to the charged Higgs boson
loops.
It is known that the scalar loop contribution to the
photonic widths is less than that of fermion and $W$ boson loops
(the last is the largest). The contributions of $W$ and $t$-quark
loops are of opposite sign, i.e., they partially compensate each
other, thus, the effect of scalar loops is enhanced here. The
coupling $\chi_{H^{\pm}}$ depends on $\lambda_5$, which is not
fixed by the observable masses. This dependence is linear:
\begin{equation}
|\chi_{\gamma\gamma}|^2 
=\frac{\Gamma_{\gamma\gamma}^{\rm 2HDM}}{\Gamma_{\gamma\gamma}^{\rm SM}}
=1-R_{\gamma\gamma}\left(1-\frac{\lambda_5}{\lambda_4}\right)
\label{Ra}
\end{equation}
with a similar dependence for the $Z\gamma$ decay.

For {\em the solutions $B$} the coupling of the charged Higgs boson $H^\pm$ 
to the observed neutral one, $\chi_{H^\pm}$, is practically independent
of $\lambda_5$ since $\chi_d+\chi_u\approx0$. 
Also, if the charged Higgs boson $H^\pm$ is heavy
(as is the case in the SM-like scenario), its coupling to the
neutral Higgs scalars $\phi$ is close to that of the vector
bosons, $\chi_{H^\pm} \approx \chi_V$. For solutions $B_{\phi\pm d}$ 
the main difference in the two-photon width from that of the SM is given
by the contribution of charged Higgs bosons, for the solution
$B_{h+u}$ also in the change of sign of the coupling to the $t$-quark as
compared to the SM case.

In Fig.~1 we show the ratio of the two-photon Higgs widths to the SM 
value for the solutions $A$ (left) and $B$ (right).
For the solutions $A$ these curves correspond to the case $\lambda_5=0$. 
Solid curves correspond to the ``exact'' case with $|\chi_i|=1$, 
for $i=q,W/Z$. 
As discussed above, these curves are below unity due to the
contribution of the charged Higgs boson, and 
for the solution $B_{h+u}$, also due
to the ``wrong'' sign of the $huu$ (or $htt$) coupling.
The shaded bands are derived from  the anticipated 1~$\sigma$ bounds
around the SM values of {\sl two} measured basic coupling constants:
for the solution $A$ for $g_d$ and $g_u$, for solution $B_{h\pm d}$
($B_{h+u}$) for $g_Z$ and $g_d$ ($g_u$).
The third basic coupling is, by virtue of the pattern equation, 
much closer to the SM value than the corresponding 
anticipated error for this quantity.

\begin{figure}[htb]
\epsfig{file=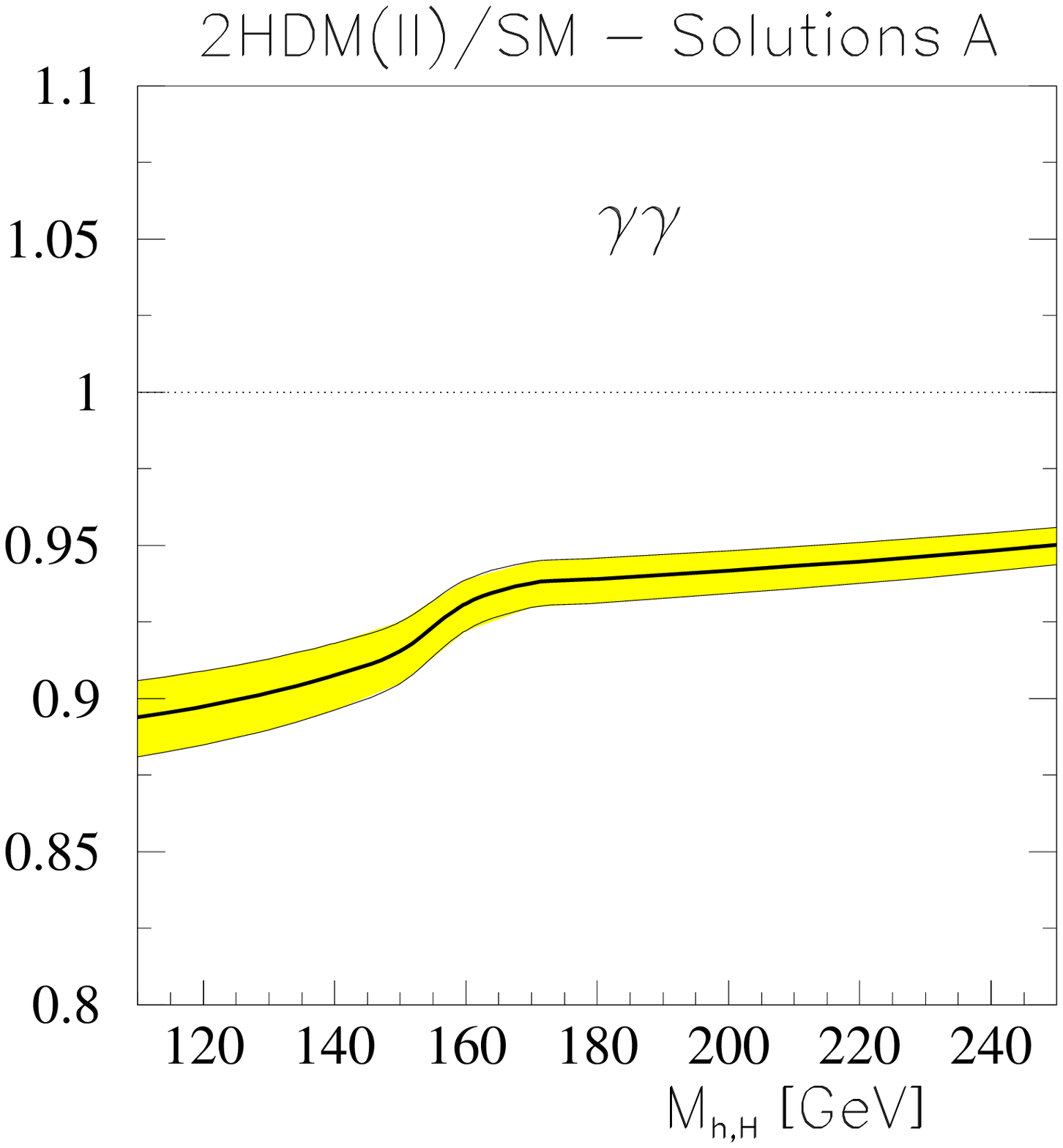,width=73mm}
\epsfig{file=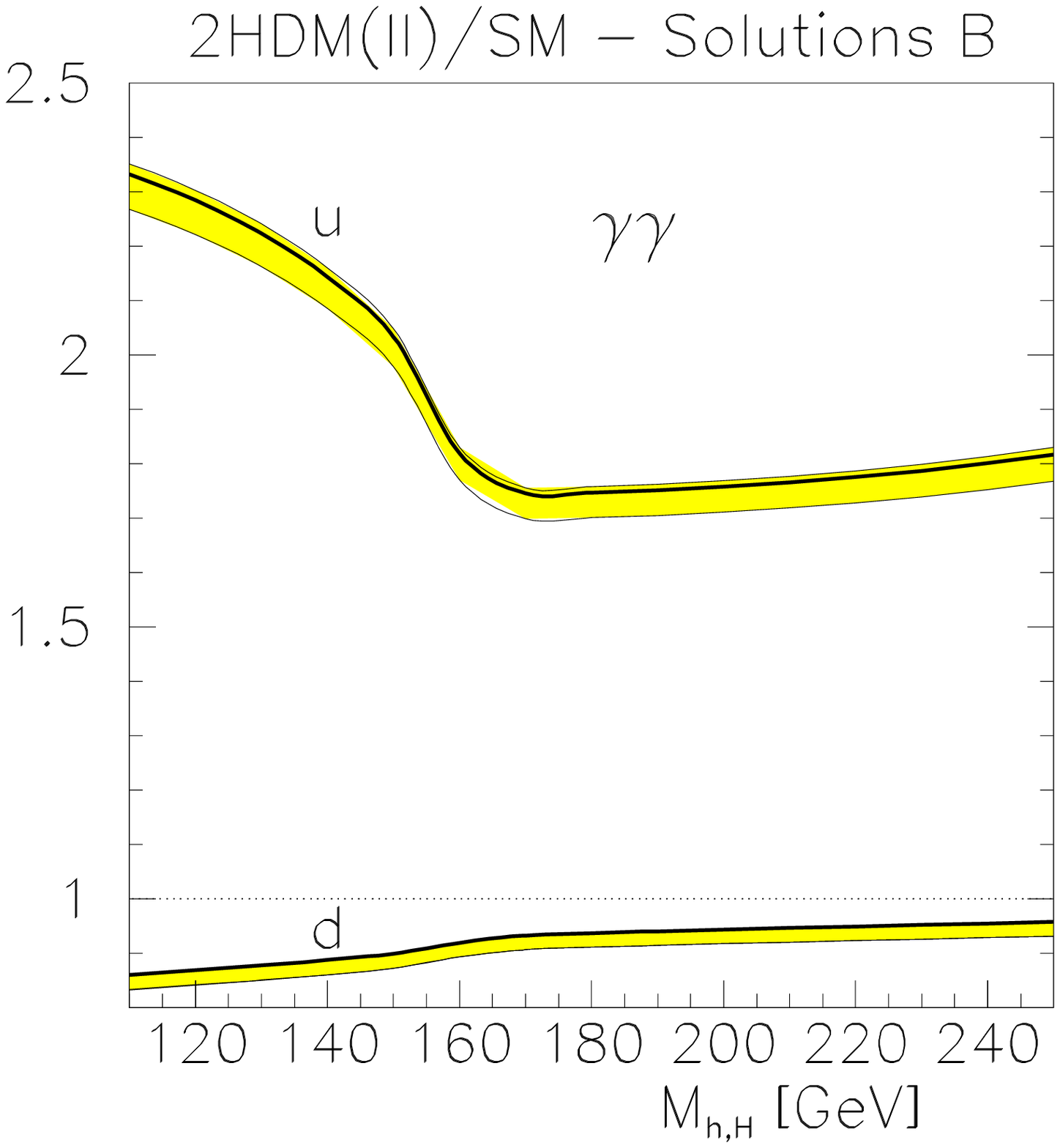,width=73mm} 
\caption{Ratios of the Higgs boson $\phi\to\gamma\gamma$ 
decay width in the 2HDM and the SM as functions of $M_{h,H}$ 
for all solutions~$A$ (with $\lambda_5=0$) and $B$.
The shaded bands
correspond to the uncertainties expected at 120~GeV.
At masses above 140~GeV, a more detailed analysis is required.}
\end{figure}
In Table~2 we present values for these ratios, and for 
comparison also for $Z\gamma$ and $gg$, for $M_\phi=120$ GeV, and
for the ``exact'' case ($|\chi_i|=1$).

\begin{table}[ht]
\begin{center}
\begin{minipage}{150mm}
\begin{tabular}{||c|c|c|c|c|c||}
\noalign{\vspace{-9pt}}
\hline\hline
solution&basic couplings&SM-like realization& $|\chi_{\gamma\gamma}|^2$ &
$|\chi_{Z\gamma}|^2$ & $|\chi_{gg}|^2$\\\hline\hline
$A_{\phi\pm}$ 
&$\chi_V\approx\chi_d\approx\chi_u\approx \pm1$ 
& $A_{h \pm}$; $A_{H-}$
&0.90 & 0.96 & 1.00\\\hline
$B_{\phi\pm d}$ 
&$\chi_V\approx-\chi_d\approx\chi_u\approx \pm1$ 
& $B_{h\pm d}$; $B_{H-d}$
&0.87 & 0.96 & 1.28 \\ \hline
$B_{\phi\pm u}$ 
& $\chi_V\approx\chi_d\approx-\chi_u\approx \pm1$
& $B_{h+u}$
&2.28 & 1.21 & 1.28 \\ \hline\hline
\end{tabular}
\end{minipage}
\end{center}
\caption{Summary of the solutions and SM-like realizations.
The last three columns give ratios of loop-induced partial widths 
to their SM values at $M_\phi=120$~GeV for the SM-like
scenarios in the 2HDM~(II) in the ``exact'' case ($|\chi_i|=1$).
The quantities for solutions $A$ depend on $\lambda_5$; 
here $\lambda_5=0$.}
\end{table}

The deviation from unity is large enough to allow a
reliable distinction of the 2HDM from the SM in the process
$\gamma\gamma\to h,H$. This conclusion is valid in a wide range of
$\lambda_5$ values. The possible precision in the determination of
$\lambda_5$ from the two-photon width depends on the
mass of the charged Higgs boson.
The deviations of the $\phi Z\gamma$ width from its SM value was 
found to be lower than that of the two photon width, see Table~2.

The two-gluon width is determined by the contributions of
$t$ and $b$ quarks. For not too high values of $\tan\beta$, the
$t$-quark contribution dominates. So, the difference 
$\chi_{gg}-1$ is determined by the difference $\chi_u-1$, and 
with high accuracy $\chi_{gg}-1 \approx 2(\chi_u-1)$.
If $\tan\beta\ll 1$ then the deviation of the Higgs boson 
coupling with $t$-quark from its SM value can for solutions $B$  
be large compared to the expected experimental uncertainty 
(\ref{couplacc}), see Table 2. 
In this case the two-gluon width can differ from its SM value 
by more than the expected experimental uncertainty (5.5\% at 
the Linear Collider), and the measurement of the two-gluon width 
could exclude the SM-like scenario from being realized by the 2HDM.
In such a case the Photon Collider can be used for a more detailed 
study of the realized non-standard Higgs sector.

\section{Conclusion}

An SM-like scenario observed  at the LHC and $e^+e^-$ Linear
Colliders can occur both in the SM and in other models, including
the 2HDM~(II). In order to distinguish these models, we implement
a pattern relation among basic couplings which is valid in the 2HDM
and in the MSSM. Taking into account anticipated uncertainties in future
measurements of the basic couplings of the Higgs boson, we found
that the considered SM-like scenario (in which partial widths of
Higgs boson decay are close to their SM values) has three types
of allowed realizations.

The comparison of the presented results for the two-photon
width with the anticipated experimental uncertainty (\ref{jik})
shows that the deviation of the two-photon width ratio from unity
is generally large enough to allow a reliable distinction of the
2HDM~(II) from the SM at a Photon Collider. For the solutions $B$
this conclusion is valid for arbitrary values of $\lambda_5$ and
arbitrary masses of other (unobserved) Higgs bosons. For the
solutions $A$, according to Eq.~(\ref{Ra}), this conclusion is
valid in a wide range of values $\lambda_5\notin (\lambda_4-
\lambda_4\delta_\gamma /R_{\gamma\gamma},\;\lambda_4+
\lambda_4\delta_\gamma /R_{\gamma\gamma})$ with
$\lambda_4=2M_{H^\pm}^2/v^2$. The possible precision in the
determination of $\lambda_5/\lambda_4$ from the two-photon width
depends strongly on the achieved precision in the determination of
the $ht\bar{t}$ coupling.

These conclusions are accurate for a Higgs boson lighter than 
140~GeV. A new analysis of experimental uncertainties is necessary
if the observed Higgs boson will be heavier than 140~GeV.
The measurement of the Higgs boson production in the process
$e\gamma\to eh$, which is sensitive to the $hZ\gamma$ coupling,
will be an additional test of the nature of the Higgs sector.

The obtained predictions are practically
identical for the cases when the observed scalar is the lighter
Higgs boson of 2HDM ($h$) or the heavier one ($H$).

\medskip

MK is grateful to the Organizing Committee for financial support.
This research has been supported by RFBR grants 99-02-17211 and
00-15-96691, and by the Research Council of Norway.

\end{document}